\documentclass[showpacs,amsmath,amssymb,prd,floatfix,preprintnumbers]{revtex4-2}
\usepackage[dvipdfmx]{graphicx}
\usepackage{bm}
\usepackage{amsmath}
\usepackage{color}
\usepackage[dvipsnames]{xcolor}
\usepackage{hyperref}
\usepackage{comment}
\usepackage{multirow}
\usepackage{url}

\hypersetup{
  bookmarksnumbered=true,
  colorlinks=true,
  linkcolor=red,
  citecolor=orange,
  urlcolor=cyan
}

\newcommand{\bk}{{\bf k}}

\newcommand{\bn}{{\bf n}}

\newcommand{\lin}{\mathrm{lin}}

\newcommand{\G}{\mathcal{G}}
\newcommand{\I}{\mathcal{I}}

\newcommand{\K}{\mathcal{K}}

\renewcommand{\L}{\mathcal{L}}
\renewcommand{\O}{\mathcal{O}}

\newcommand{\bin}{\mathrm{bin}}

\newcommand{\simgt}{\lower.5ex\hbox{$\; \buildrel > \over \sim \;$}}
\newcommand{\simlt}{\lower.5ex\hbox{$\; \buildrel < \over \sim \;$}}

\begin{document}

\title[]{Fast computation of the non-Gaussian covariance of redshift-space galaxy power spectrum multipoles}

\author{Yosuke~Kobayashi${}^{1}$}
\email{yosukekobayashi@arizona.edu}
\affiliation{
${}^{1}$ Department of Astronomy/Steward Observatory, University of Arizona, 933 North Cherry Avenue, Tucson, Arizona 85721, USA
}

\begin{abstract}
The non-Gaussian part of the covariance matrix of the galaxy power spectrum involves the connected four-point correlation in Fourier space, i.e., trispectrum.
This paper introduces a fast method to compute the non-Gaussian part of the covariance matrix of the galaxy power spectrum multipoles in redshift space at tree-level standard perturbation theory. 
For the tree-level galaxy trispectrum, the angular integral between two wavevectors can be evaluated analytically by employing an FFTLog.
The new implementation computes the non-Gaussian covariance of the power spectrum monopole, quadrupole, hexadecapole and their cross-covariance in $\O(10)$ seconds, for an effectively arbitrary number of instances of cosmological and galaxy bias parameters and redshift, without any parallelization or acceleration.
It is a large advantage over conventional numerical integration.
We demonstrate that the computation of the covariance at $k = 0.005-0.4 \, h \, \mathrm{Mpc}^{-1}$ gives results with $0.1-1\%$ accuracy.
The efficient computation of the analytic covariance can be useful for future galaxy surveys, especially utilizing multi-tracer analysis.
\end{abstract}

\maketitle

\section{Introduction}
\label{sec:intro}

Galaxy clustering, i.e., the spatial distributions of galaxies on the large-scale structure of the Universe, is an important cosmological probe.
Spectroscopic galaxy surveys are an indispensable tool for accessing the cosmological information contained in galaxy clustering statistics, the most important of which is the two-point correlation function or its Fourier counterpart, the power spectrum.
In recent years, so-called full-shape analyses of the galaxy power spectrum (e.g. Refs.~\cite{DAmico:2019fhj,Ivanov:2019pdj,Chen:2021wdi,Philcox:2021kcw,Kobayashi:2021oud}) have been developed and put tight constraints on the parameters of the standard $\Lambda$CDM cosmology.
This type of analysis is expected to become the main goal of ongoing and future galaxy surveys, such as Dark Energy Spectroscopic Instrument (DESI; \cite{DESI:2016fyo}), \textit{Euclid} satellite \cite{EUCLID:2011zbd}, SPHEREx satellite \cite{Dore:2014cca}, Subaru Prime Focus Spectrograph (PFS; \cite{PFSTeam:2012fqu}), and Roman Space Telescope \cite{Gehrels:2014spa}.

Cosmology inferences from the galaxy clustering data of current spectroscopic surveys usually assume a Gaussian likelihood, where the statistical uncertainty in the clustering signals follows the Gaussian distribution determined by a covariance matrix.
Except on very large scales where we can sample only a small number of modes, this assumption is a good approximation of the likelihood of galaxy clustering statistics.
Hence, accurate models for the covariance are crucial for obtaining accurate parameter constraints.
In current analyses of spectroscopic surveys, the covariance matrix is typically estimated from the mock galaxy catalogs created by large simulation ensembles.
Simulation-based mock catalogs have the advantage that they can incorporate the highly nonlinear structure of the distributions of target galaxies and the nontrivial survey masks.
On the other hand, it is computationally costly to produce thousands of realizations of mock catalogs from full $N$-body simulations with sufficient volume and resolution.
Thus, the approximated simulation approaches for a massive mock generation have been devised, e.g. Refs.~\cite{Chuang:2014vfa,Kitaura:2015uqa,Rodriguez-Torres:2015vqa,Balaguera-Antolinez:2022xko}.
Further, the covariance estimated from simulations has statistical noise due to the finite number of realizations, and the correction to it increases the errors of the inferred cosmological parameters \cite{Hartlap:2006kj,Dodelson:2013uaa,Percival:2013sga}. 

Analytic modeling of the covariance matrix based on perturbation theory (PT) is a complementary approach that has been pursued in e.g. Refs.~\cite{Meiksin:1998mu,Scoccimarro:1999kp,Smith:2008ut,Chan:2016ehg,Sugiyama:2019ike,Wadekar:2019rdu}.
These works show that the PT-based approach successfully predicts the power spectrum covariance, including the off-diagonal components arising from the non-Gaussian nature of galaxy distributions.
The main advantage of the analytic approach is two-fold: one is that there is no sampling noise arising from a finite number of simulation realizations that leads to the degradation of parameter constraints, and another is the relatively cheap computational cost.

Recently, Ref.~\cite{Wadekar:2019rdu} showed a model of the PT-based analytic covariance of the galaxy power spectrum multipoles in redshift space, which incorporated various contributions that are physical or observational in a realistic galaxy survey situation.
Also, Ref.~\cite{Wadekar:2020hax} showed that this covariance model results in the cosmological constraints very similar to those obtained using the simulation mock-based covariance matrix by Refs.~\cite{Kitaura:2015uqa,Rodriguez-Torres:2015vqa}, for the galaxy power spectrum of the Sloan Digital Sky Survey (SDSS)-III Baryon Oscillation Spectroscopic Survey (BOSS; \cite{BOSS:2012dmf}).
While their analytic covariance model is much faster to compute than mocks for simulation-based covariances, the non-Gaussian part from the connected four-point correlation of galaxy density fields, i.e., the trispectrum is still a computational bottleneck.
Depending on the number of wavenumber bins, it can take several minutes, or more, for each input cosmology and galaxy bias parameter.
The computation time for covariance matrices for multiple combinations of multipoles, at multiple redshift bins, and for multi-tracer analyses can be substantial.

The so-called FFTLog \cite{TALMAN197835,Hamilton:1999uv}, which denotes the fast Fourier transform (FFT) on the data sampled at logarithmically spaced points, can be used to compute certain integrals much more efficiently than direct numerical integration.
In the past several years, several works have applied FFTLog techniques to cosmological computations \cite{Schmittfull:2016jsw,Schmittfull:2016yqx,McEwen:2016fjn,Fang:2016wcf,Assassi:2017lea,Simonovic:2017mhp,Schoneberg:2018fis,Fang:2019xat,Lee:2020ebj,Fang:2020vhc}.
Many of these works follow a common base strategy: first, they decompose the integrand into functions whose integrals can be evaluated analytically.
In this step, they operate the FFTLog on some cosmology-dependent quantity, typically the linear power spectrum, to decompose it into a linear combination of power-law functions.
Second, they substitute the analytic solution into the integral of each component of the decomposed integrand and collect them to obtain the desired integral.
This base strategy is highly versatile, and we adopt it in this work to compute the non-Gaussian covariance coming from the trispectrum contribution, as described in Sec.~\ref{sec:method}.
Our method can compute covariance matrices, for an effectively arbitrary number of combinations of multipoles/redshifts/cosmology and galaxy bias parameters, within $\O(10)$ seconds.

This paper is organized as follows: in Sec.~\ref{sec:non_gauss_cov}, we make a basic description of the non-Gaussian covariance coming from the galaxy trispectrum and its modeling based on the tree-level PT.
Section~\ref{sec:method} describes the computationally expensive part of the non-Gaussian covariance and suggests an FFTLog-based technique to compute that part efficiently.
Section~\ref{sec:results} shows the demonstration of the method and discusses the computation accuracy and time. 
Finally, we summarize this work and make a conclusion in Sec.~\ref{sec:conclusion}.

\section{Non-Gaussian power spectrum covariance from galaxy trispectrum}
\label{sec:non_gauss_cov}

In this section, we describe the non-Gaussian part of the covariance matrix coming from the ``regular'' trispectrum \cite{Wadekar:2019rdu}, which has parallelogram-shaped momentum configurations.
The super-sample covariance, which arises from the beat modes of the trispectrum \cite{Hamilton:2005dx,Takada:2013wfa,Wadekar:2019rdu}, needs to be evaluated separately.
In a survey volume $V$, the covariance contribution from the regular trispectrum is
\begin{align}
\label{eq:cov_T0}
    C^{T_0}_{\ell_1,\ell_2}(k_1,k_2) 
    &= \frac{1}{V} \frac{1}{N_{k_1}} \sum_{\bk'_1 \in \bin~1} \frac{1}{N_{k_2}} \sum_{\bk'_2 \in \bin~2} (2\ell_1+1) \L_{\ell_1}(\hat{\bk}'_1 \cdot \hat{\bn}) (2\ell_2+1) \L_{\ell_2}(\hat{\bk}'_2 \cdot \hat{\bn}) T(\bk'_1,-\bk'_1,\bk'_2,-\bk'_2) \nonumber\\
    &\approx \frac{1}{V} \int_{\bk'_1 \in \bin~1} \frac{\mathrm{d}^3k'_1}{V_{k_1}} \int_{\bk'_2 \in \bin~2} \frac{\mathrm{d}^3k'_2}{V_{k_2}} (2\ell_1+1) \L_{\ell_1}(\hat{\bk}'_1 \cdot \hat{\bn}) (2\ell_2+1) \L_{\ell_2}(\hat{\bk}'_2 \cdot \hat{\bn}) T(\bk'_1,-\bk'_1,\bk'_2,-\bk'_2),
\end{align}
where $\L_\ell(\cdot)$ is the $\ell$th-order Legendre polynomial, $\hat{\bk}'_i$ and $\hat{\bn}$ are the unit vectors along the wavevector $\bk'_i$ and along the line-of-sight direction, respectively.
$N_{k_i}$ is the number of Fourier modes that enter the $i$th bin of $k$, and $V_{k_i}$ is the volume of the Fourier-space shell in which the modes belong to the $i$th bin.
We can see Eq.~(\ref{eq:cov_T0}) as the bin-averaged trispectrum, with the weights of Legendre polynomials.
However, we do not consider the effect of bin averaging in this paper, expecting that the trispectrum does not vary drastically in each $k$-bin.
In this thin-shell approximation, the integrals along the radial directions of $\bk'_1$ and $\bk'_2$ are dropped, and Eq.~(\ref{eq:cov_T0}) reduces to the angular integral over two solid angles $\Omega_{\hat{\bk}_1}$ and $\Omega_{\hat{\bk}_2}$,
\begin{align}
\label{eq:cov_T0_sa_int}
    C^{T_0}_{\ell_1 \ell_2}(k_1,k_2) &\approx \frac{1}{V} \int \frac{\mathrm{d}^2 \Omega_{\hat{\bk}_1}}{4\pi} \int \frac{\mathrm{d}^2 \Omega_{\hat{\bk}_2}}{4\pi} (2\ell_1+1) \L_{\ell_1}(\hat{\bk}_1 \cdot \hat{\bn}) (2\ell_2+1) \L_{\ell_2}(\hat{\bk}_2 \cdot \hat{\bn}) T(\bk_1,-\bk_1,\bk_2,-\bk_2).
\end{align}

Following Ref.~\cite{Wadekar:2019rdu}, we use tree-level standard perturbation theory (SPT; see Ref.~\cite{Bernardeau:2001qr} for a review) to model the galaxy trispectrum in Eq.~(\ref{eq:cov_T0}).
The tree-level galaxy trispectrum is 6th order in linear density perturbations; assuming Gaussian initial conditions, and it can be written as \cite{Fry:1983cj}
\begin{align}
\label{eq:trispec_tree}
    T(\bk_1,\bk_2,\bk_3,\bk_4) &= T_{2211}(\bk_1,\bk_2,\bk_3,\bk_4) + T_{3111}(\bk_1,\bk_2,\bk_3,\bk_4),
\end{align}
where
\begin{align}
    T_{2211}(\bk_1,\bk_2,\bk_3,\bk_4) &= 4 Z_1(\bk_1) Z_1(\bk_2) Z_2(-\bk_1,-\bk_1-\bk_3) Z_2(\bk_2,\bk_1+\bk_3) P_{\lin}(k_1) P_{\lin}(k_2) P_{\lin}(|\bk_1+\bk_3|) + \text{(11 perms.)}, \\
    T_{3111}(\bk_1,\bk_2,\bk_3,\bk_4) &= 6 Z_1(\bk_1) Z_1(\bk_2) Z_1(\bk_3) Z_3(\bk_1,\bk_2,\bk_3) P_{\lin}(k_1) P_{\lin}(k_2) P_{\lin}(k_3) + \text{(3 perms.)}.
\end{align}
Here, $T_{2211}$ (also called the ``snake'' term) is the correlation among two linear-order perturbations and two second-order ones, whereas $T_{3111}$ (the ``star'' term) is the correlation among three linear-order perturbations and a third-order one.
$P_{\lin}$ is the linear matter power spectrum.
We give the explicit formulas of the redshift-space PT kernels $Z_n$ in Appendix~\ref{sec:pt_kernel}.

For the parallelogram configurations contributing to non-Gaussian covariance Eq.~(\ref{eq:cov_T0_sa_int}), the trispectrum reduces to
\begin{align}
    T(\bk_1,-\bk_1,\bk_2,-\bk_2)
    &= \left[ 8 P_{\lin}^2(k_1) P_{\lin}(|\bk_1+\bk_2|) Z_1^2(\bk_1) Z_2^2(-\bk_1,\bk_1+\bk_2) + (\bk_1 \leftrightarrow \bk_2) \right] \nonumber\\
    & \,\, + 16 P_{\lin}(k_1) P_{\lin}(k_2) P_{\lin}(|\bk_1+\bk_2|) Z_1(\bk_1) Z_1(\bk_2) Z_2(-\bk_1,\bk_1+\bk_2) Z_2(-\bk_2,\bk_1+\bk_2).
\end{align}
Substituting the above into Eq.~(\ref{eq:cov_T0_sa_int}) yields
\begin{align}
\label{eq:cov_T0_detail}
    C^{T_0}_{\ell_1,\ell_2}(k_1,k_2) &= \frac{1}{V} \int_{\hat{\bk}_{\ell_1}, \hat{\bk}_{\ell_2}} 8 P_{\lin}(|\bk_1+\bk_2|) \left[ P_{\lin}^2(k_1) Z_1^2(\bk_1) Z_2^2(-\bk_1,\bk_1+\bk_2) + (\bk_1 \leftrightarrow \bk_2) \right] \nonumber\\
    &\,\, + \frac{1}{V} \int_{\hat{\bk}_{\ell_1}, \hat{\bk}_{\ell_2}} 16 P_{\lin}(|\bk_1+\bk_2|) P_{\lin}(k_1) P_{\lin}(k_2) Z_1(\bk_1) Z_1(\bk_2) Z_2(-\bk_1,\bk_1+\bk_2) Z_2(-\bk_2,\bk_1+\bk_2) \nonumber\\
    &\,\, + \frac{1}{V} \int_{\hat{\bk}_{\ell_1}, \hat{\bk}_{\ell_2}} 12 \left[ P_{\lin}^2(k_1) P_{\lin}(k_2) Z_1^2(\bk_1) Z_1(\bk_2) Z_3(\bk_1,-\bk_1,\bk_2) + (\bk_1 \leftrightarrow \bk_2) \right],
\end{align}
where we define the notation for the shell integral
\begin{align}
\label{eq:brev_notation}
    \int_{\hat{\bk}_{\ell_i}} \rightarrow \int \frac{\mathrm{d}^2 \Omega_{\hat{\bk}_i}}{4\pi} (2\ell_i+1) \L_{\ell_i}(\hat{\bk}_i \cdot \hat{\bn}),
\end{align}
for brevity.
Equation~(\ref{eq:cov_T0_detail}) is the formula that we aim to efficiently compute in this work.
In the above formulation, we ignore the nontrivial survey window in actual observations. 
However, one can obtain the expression taking into account the survey window, Eq.~(66) of Ref.~\cite{Wadekar:2019rdu}, by replacing a factor $1/V$ in Eq.~(\ref{eq:cov_T0_detail}) as $I_{44}/I_{22}^2$ defined in Ref.~\cite{Wadekar:2019rdu}.

\section{FFTLog-based computation of non-Gaussian covariance}
\label{sec:method}

Equation~(\ref{eq:cov_T0_sa_int}) further reduces to a one-dimensional integral with respect to the angle between $\bk_1$ and $\bk_2$, i.e.,
\begin{align}
\label{eq:cov_T0_int_reduction}
    C^{T_0}_{\ell_1 \ell_2}(k_1,k_2) &= \frac{1}{V} \int_{-1}^{1} \frac{\mathrm{d}(\hat{\bk}_1 \cdot \hat{\bk}_2)}{2} T_{\ell_1 \ell_2}(k_1, k_2, \hat{\bk}_1 \cdot \hat{\bk}_2),
\end{align}
where we introduce $T_{\ell_1 \ell_2}(k_1, k_2, \hat{\bk}_1 \cdot \hat{\bk}_2)$ to denote the trispectrum integrand after the reduction of the two solid-angle integrals to a single integral with respect to $\hat{\bk}_1 \cdot \hat{\bk}_2$, i.e., the cosine of the angle between $\hat{\bk}_1$ and $\hat{\bk}_2$.
In this form, the integral of the $T_{3111}$ term of Eq.~(\ref{eq:cov_T0_detail}) can be done.
The other terms coming from the $T_{2211}$ cannot be integrated analytically, due to $P_{\lin}(|\bk_1+\bk_2|)$ terms, and cannot be simplified further.
While it is a one-dimensional integral, the integrand can be steeply varying and is confined to a narrow region in the entire integration range, especially when two bins $k_1$ and $k_2$ are similar in magnitude.
Thus, it requires an adaptive quadrature method to achieve sufficient precision.

In this paper, we use an FFTLog method \cite{McEwen:2016fjn,Simonovic:2017mhp} to approximate the linear power spectrum by a linear combination of the power-law functions of $k$ with complex exponents:
\begin{align}
\label{eq:decomp_pk_lin}
    \bar{P}_\lin(k) = \sum_{m=-N/2}^{N/2} c_m k^{\nu+i\eta_m},
\end{align} 
where the coefficients $c_m$ and exponents $\eta_m$ are given by
\begin{align}
\label{eq:coef_power}
    c_m = W_m k^{-\nu-i\eta_m} \frac{1}{N} \sum_{j=0}^{N-1} P_\lin(k_j) \left( \frac{k_j}{k_\mathrm{min}} \right)^{-\nu} e^{-2\pi i mj / N}, \,\, \eta_m = \frac{N-1}{N} \frac{2\pi m}{\ln (k_\mathrm{max}/k_\mathrm{min})}.
\end{align}
This decomposition is done by the FFTLog \cite{TALMAN197835,Hamilton:1999uv}, i.e., performing a one-dimensional FFT on the linear power spectrum sampled at $N$ logarithmically spaced points of $k$ in the range $[k_\mathrm{min}$, $k_\mathrm{max}]$.
The parameter $\nu$ in Eq.~(\ref{eq:decomp_pk_lin}) is commonly referred to as ``bias'' in the literature.
We refer to it as ``FFT power-law bias'', to avoid confusion with the galaxy bias parameters.
The weights in $c_m$ are $W_m = 1/2$ at the endpoints $(m = -N/2, N/2)$, and $W_m = 1$ otherwise. 

The integral Eq.~(\ref{eq:cov_T0_int_reduction}) coming from the $T_{2211}$ term, which we denote as $C^{T_{2211}}_{\ell_1,\ell_2}$, can be decomposed into a linear combination of more elementary integrals:
\begin{align}
\label{eq:cov_decomp}
    C^{T_{2211}}_{\ell_1,\ell_2}(k_1,k_2) &= \frac{1}{V} \sum_{\alpha,\beta} f_{\ell_1,\ell_2}^{\alpha,\beta}(k_1,k_2) \I_{\alpha,\beta}(k_1,k_2),
\end{align}
where we define the elementary integrals as
\begin{align}
\label{eq:integral_element}
    \I_{\alpha,\beta}(k_1,k_2) &\equiv \int_{-1}^{1} \frac{\mathrm{d}(\hat{\bk}_1 \cdot \hat{\bk}_2)}{2} \frac{(\hat{\bk}_1 \cdot \hat{\bk}_2)^{\beta}}{|\bk_1+\bk_2|^{\alpha}} P_{\lin}(|\bk_1+\bk_2|).
\end{align}
Using the power-law decomposition of the linear power spectrum Eq.~(\ref{eq:decomp_pk_lin}), we can further decompose the integral Eq.~(\ref{eq:integral_element}) as
\begin{align}
\label{eq:base_master_relation}
    \I_{\alpha,\beta}(k_1,k_2) 
    &= \sum_{m} c_m \int_{-1}^{1} \frac{\mathrm{d}(\hat{\bk}_1 \cdot \hat{\bk}_2)}{2} \frac{(\hat{\bk}_1 \cdot \hat{\bk}_2)^{\beta}}{|\bk_1+\bk_2|^{\alpha-\nu - i\eta_m}} 
    = \sum_{m} c_m \mathsf{I}_{\alpha_m, \beta}(k_1,k_2).
\end{align}
In the above, we define a complex-valued index $\alpha_m \equiv \frac{1}{2} (\alpha-\nu-i\eta_m)$ and the integral that has an analytic solution: 
\begin{align}
\label{eq:master_integral}
    \mathsf{I}_{a,b}(k_1,k_2) &\equiv \int_{-1}^{1} \frac{\mathrm{d}\mu}{2} \frac{\mu^{b}}{(k_1^2 + k_2^2 + 2 k_1 k_2 \mu)^{a}} \nonumber\\
    &= \frac{(k_1^2+k_2^2)^{-a}}{2(b+1)} \left\{ {_2F_1}\left(a, b+1; b+2; -\frac{2k_1 k_2}{k_1^2+k_2^2} \right) + (-1)^{b} {_2F_1}\left(a, b+1; b+2; \frac{2k_1 k_2}{k_1^2+k_2^2} \right) \right\},
\end{align}
where ${_2F_1}$ is the Gauss hypergeometric function.
Note that in our problem, $a$ is a complex number and $b$ is a non-negative integer.
We refer to this formula as the ``master integral,'' as it is the building block of our method.
The implementation of the master integral is described in Appendix~\ref{sec:master_integral}.

Finally, by combining Eqs.~(\ref{eq:cov_decomp}) and (\ref{eq:base_master_relation}), the computationally expensive covariance term is expressed by
\begin{align}
    \label{eq:cov_decomp_final}
    C^{T_{2211}}_{\ell_1,\ell_2}(k_1,k_2) &= \frac{1}{V} \sum_{m,\alpha,\beta} c_m f_{\ell_1,\ell_2}^{\alpha,\beta}(k_1,k_2) \mathsf{I}_{\alpha_m, \beta}(k_1,k_2),
\end{align}
which is a linear combination of the master integrals Eq.~(\ref{eq:master_integral}), and there is no need to use numerical integration if we can evaluate the master integrals quickly enough.

In summary, to implement the non-Gaussian covariance, we have the following procedures:
\begin{itemize}
    \item[(1)] decompose the linear power spectrum using the FFTLog to obtain the coefficients $c_m$ and the exponents $\eta_m$.
    $c_m$ contains all the cosmology dependence of the linear power spectrum, while $\eta_m$ is determined purely by the FFTLog setting, i.e., $\{k_\mathrm{min}, k_\mathrm{max}, N\}$ [see Eq.~(\ref{eq:coef_power})].
    \item[(2)] decompose the integrals in Eq.~(\ref{eq:cov_T0_detail}) into the linear combinations of the element integrals Eq.~(\ref{eq:integral_element}), and identify the coefficient functions $f_{\ell_1,\ell_2}^{\alpha,\beta}(k_1,k_2)$ for all combinations of the power indices $(\alpha, \beta)$, given the multipoles $\{\ell_1, \ell_2\}$.
    In Table~\ref{tab:combination_ab}, we list the $(\alpha, \beta)$ combinations necessary to compute the covariance matrices for different sets of multipoles.
    \item[(3)] combine the above two with the master integral Eq.~(\ref{eq:master_integral}) to compute the final expression Eq.~(\ref{eq:cov_decomp_final}). 
\end{itemize}
There are two points to note: 
first, the integral $\mathsf{I}_{\alpha_m, \beta}(k_1,k_2)$ in Eq.~(\ref{eq:cov_decomp_final}) does not depend on cosmology, redshift, and galaxy bias parameters, and hence we can precompute it once we have fixed the $k$ bins and the FFTLog setting.
Second, so far we have focused on the non-Gaussian covariance without shot noise, but the shot-noise contribution can also be computed using our method.
We show the result of the shot-noise contribution in Appendix~\ref{sec:shotnoise}.

\begin{table}
\begin{center}
\begin{tabular}{ll} \hline \hline
    Multipoles & $(\alpha, \beta)$ \\ 
    \hline
    $\ell = \{ 0 \}$ & (4, 0), (4, 1), (4, 2), (4, 3), (4, 4), (4, 5), (4, 6), (4, 7), (4, 8) \\
    $\ell = \{ 0, 2 \}$ & (4, 0), (4, 1), (4, 2), (4, 3), (4, 4), (4, 5), (4, 6), (4, 7), (4, 8), (4, 9), (4, 10) \\
    $\ell = \{ 0, 2, 4 \}$ & (0, 0), (0, 2), (0, 4), (2, 0), (2, 1), (2, 2), (2, 3), (2, 4), (2, 5), (2, 6), (4, 0), \\ 
    & (4, 1), (4, 2), (4, 3), (4, 4), (4, 5), (4, 6), (4, 7), (4, 8), (4, 9), (4, 10), (4, 11), (4, 12) \\
    \hline \hline
\end{tabular}
\caption{
The combinations of $(\alpha, \beta)$ needed to compute Eq.~(\ref{eq:cov_decomp}) for each set of the multipole moments.
To compute the covariance of the monopole moment only, we need to consider nine combinations of $(\alpha, \beta)$ listed corresponding to $\ell = \{0\}$.
Likewise, the covariance of the monopole, quadrupole (and hexadecapole) and their cross-covariance requires 11 (23) combinations corresponding to $\ell = \{0,2\}$ ($\ell = \{0,2,4\}$).
}
\label{tab:combination_ab}
\end{center}
\end{table}

\section{Numerical Results}
\label{sec:results}

In this section, we demonstrate the method derived in the previous section and compare its result to the numerical integration.
We implement the $C^{T_{2211}}$ computation using \texttt{numpy.fft.fft} for the power-law decomposition of the linear power spectrum, and \texttt{scipy.integrate.quad} for direct numerical integration of the trispectrum integrand of the form Eq.~(\ref{eq:cov_T0_int_reduction}).
The trispectrum integrand is derived by modifying the \textit{Mathematica} notebook in the publicly available \textsc{CovaPT} \footnote{\url{https://github.com/JayWadekar/CovaPT}} code \cite{Wadekar:2019rdu}.

\subsection{Computation accuracy}

We compute the linear matter power spectrum at $z = 0$ with the publicly available Boltzmann code \textsc{CAMB} \cite{Lewis:2002ah}.
The input cosmology is the flat $\Lambda$CDM model specified by \textit{Planck} 2015 best-fit \cite{Planck:2015fie}: $\omega_\mathrm{b} = 0.02225$, $\omega_\mathrm{cdm} = 0.1198$, $\Omega_\mathrm{m} = 0.3156$, $\ln(10^{10} A_\mathrm{s}) = 3.094$, and $n_\mathrm{s} = 0.9645$, where $\omega_\mathrm{b}$ and $\omega_\mathrm{cdm}$ are the physical density parameters of baryons and cold dark matter, respectively, $\Omega_\mathrm{m}$ is the total matter density parameter, $A_\mathrm{s}$ is the amplitude of the primordial curvature power spectrum, and $n_\mathrm{s}$ is its spectral tilt at the pivot scale $k_\mathrm{pivot} = 0.05\,\mathrm{Mpc}^{-1}$.
We set the massive neutrino density $\omega_{\nu} = 0.00064$.
To decompose the linear power spectrum, we apply the FFTLog operation with $\{k_\mathrm{min}, k_\mathrm{max}\} = \{10^{-5}, 10\} \, h\,\mathrm{Mpc}^{-1}$ and $N = 512$. We set the FFT power-law bias $\nu = -0.3$.

Under this setting, we compute the master integrals for the linearly spaced $k$-bins centered in the range $[0.005,0.395]$ $h\,\mathrm{Mpc}^{-1}$ with the bin width $\Delta k = 0.01\,h\,\mathrm{Mpc}^{-1}$.
This corresponds to a covariance matrix with size ($40 \times 40$) for each combination of multipole moments $\{\ell_1, \ell_2\}$.
By combining the precomputed master integrals with the decomposed linear power spectrum and the coefficient functions $f_{\ell_1,\ell_2}^{\alpha,\beta}(k_1,k_2)$, we compute the covariance matrices for all the combinations of the power spectrum monopole, quadrupole, and hexadecapole.
The coefficient functions are derived from the first- and second-order PT kernels $Z_1$ and $Z_2$, and thus they depend on the linear growth rate $f \equiv \mathrm{d}\ln D / \mathrm{d}\ln a$ and the galaxy bias parameters $\{ b_1,b_2,b_{\G_2} \}$ (see Appendix~\ref{sec:pt_kernel} for detail).
We set $f = 0.528$ following the $\Lambda$CDM cosmology assumed above, and $b_1 = 2, b_2 = -1, b_{\G_2} = 0.1$ as an example.
Since the $C^{T_{3111}}$, which can be evaluated analytically, is not our focus, we set the bias parameters that appear starting from the third order, i.e., $\{ b_3, b_{\G_3}, b_{\delta \G_2}, b_{\Gamma_3} \}$, to be zero.
Note that we can also obtain the covariance in real space (i.e., not in redshift space) by setting $f = 0$.
The survey volume is set to $V = 1\,(h^{-1}\,\mathrm{Gpc})^{3}$.

\begin{figure}
\centering
  \begin{tabular}{c}
    \begin{minipage}{\hsize}
    \centering
    \includegraphics[width=1\columnwidth]{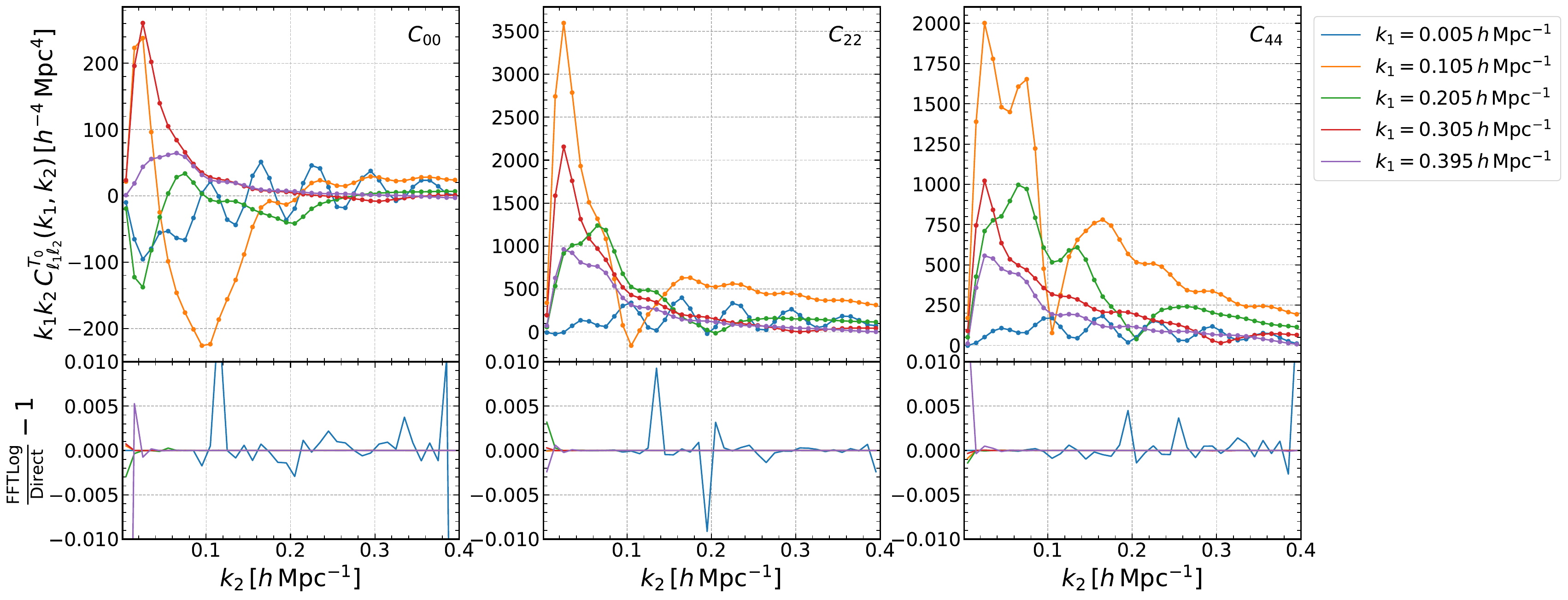}
    \end{minipage} \\
    
    \begin{minipage}{\hsize}
    \centering
    \includegraphics[width=1\columnwidth]{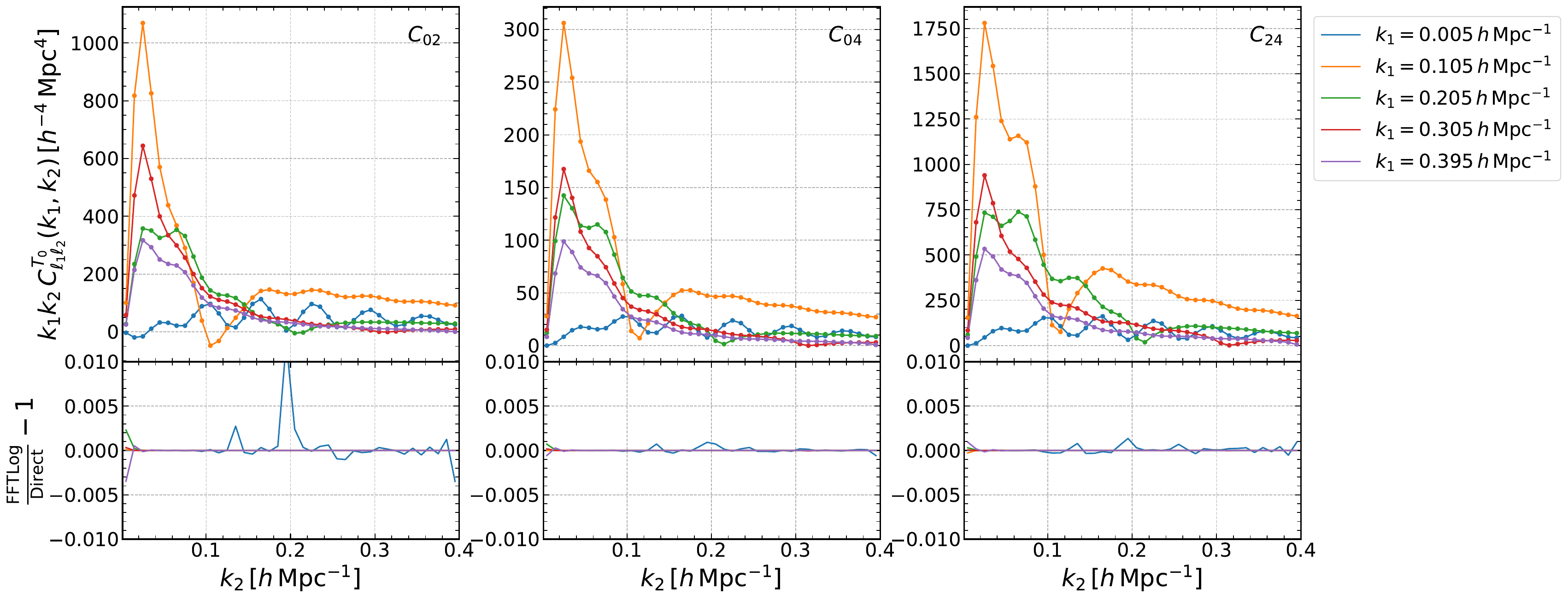}
    \end{minipage}
  \end{tabular}
\caption{
Comparison of the non-Gaussian covariance matrix coming from the whole regular trispectrum $(= T_{2211} + T_{3111})$, between the FFTLog-based method proposed in this paper and direct numerical integration.
Each of the six blocks shows a combination of multipole moments.
For each block, the upper panel shows the non-Gaussian covariance computed with the FFTLog-based method (solid lines) and with direct integration (filled symbols), and the lower panel shows the fractional difference between the two.
In all cases, the FFTLog-based method agrees with direct integration well within $0.1-1\%$. 
}
\label{fig:comparison_fftlog_vs_direct_kmax0.4}
\end{figure}

\begin{figure}
\centering
  \begin{tabular}{c}
    \begin{minipage}{\hsize}
    \centering
    \includegraphics[width=1\columnwidth]{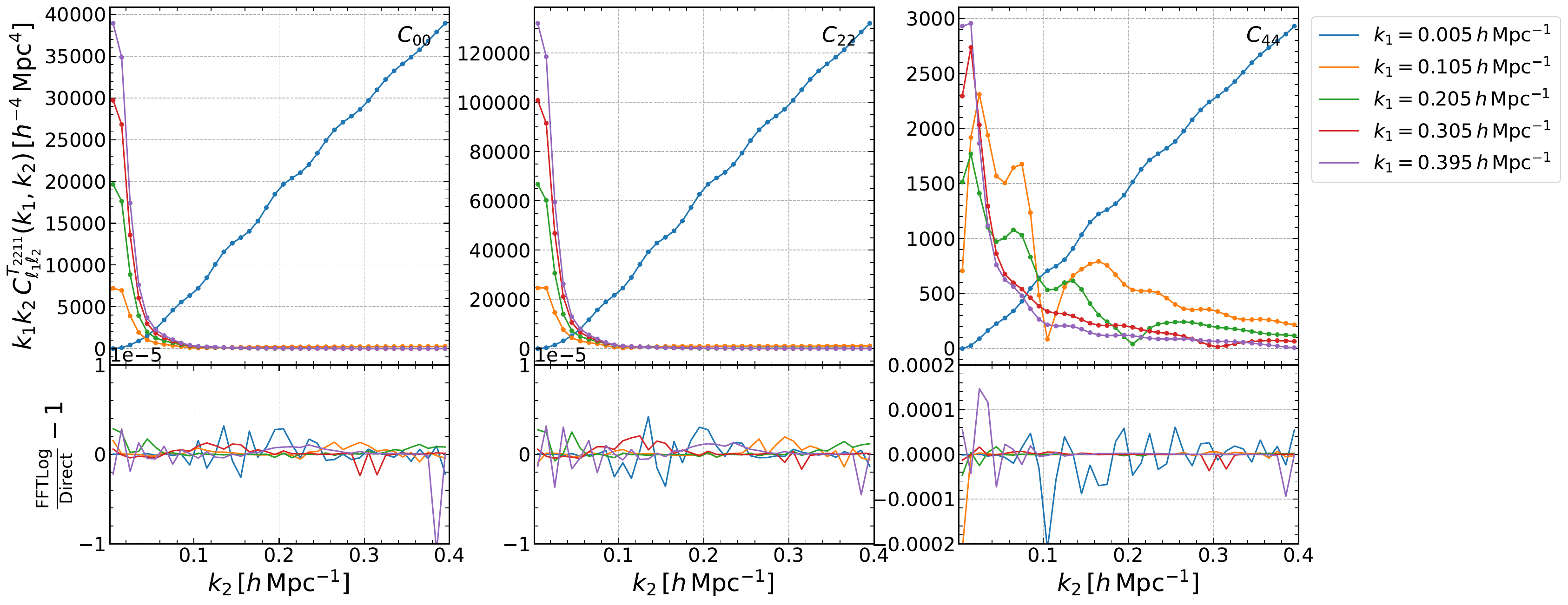}
    \end{minipage} \\
    
    \begin{minipage}{\hsize}
    \centering
    \includegraphics[width=1\columnwidth]{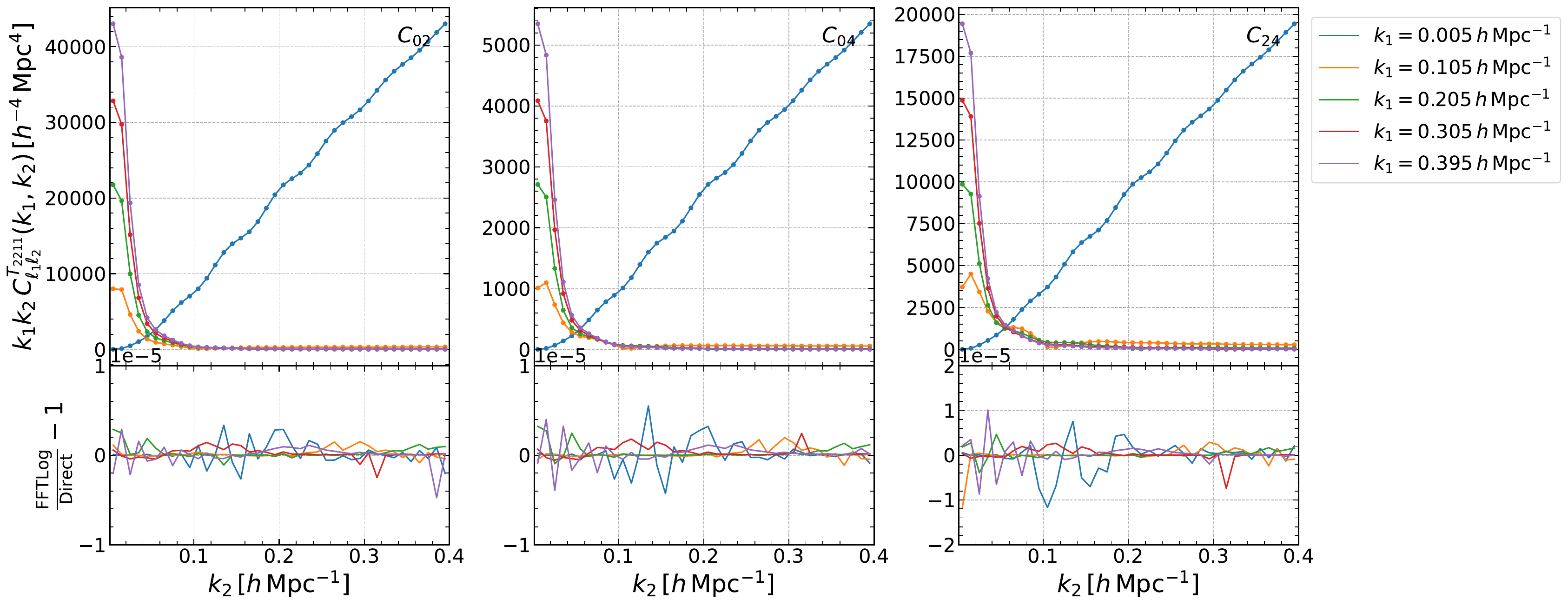}
    \end{minipage}
  \end{tabular}
\caption{
Comparison of $C^{T_{2211}}$ term of the covariance between the FFTLog-based method and direct numerical integration.
In all cases except for $C_{44}$, the auto-covariance of the hexadecapole, the FFTLog-based method agrees with direct integration within $0.001\%$.
The computation accuracy is relatively worse for $C_{44}$, but still around $0.01\%$.
}
\label{fig:comparison_T2211_kmax0.4}
\end{figure}

Figure~\ref{fig:comparison_fftlog_vs_direct_kmax0.4} shows a comparison of the non-Gaussian covariance $C^{T_0}_{\ell_1 \ell_2}(k_1,k_2)$ between our method and the conventional direct numerical integration, for all six combinations of power spectrum multipoles $\ell = \{0,2,4\}$.
For each multipole combination, the upper panel shows the trispectrum part of the covariance we compute using our FFTLog-based method (solid lines) and direct numerical integration (filled symbols).
Note that it shows the whole trispectrum contribution, $C^{T_{2211}} + C^{T_{3111}}$, while we apply our method only to $C^{T_{2211}}$ term.
We see that all cases show good agreement between the FFTLog-based method and direct integration, within $1\%$, reproducing the complicated structure of off-diagonal covariance.
The numerical accuracy is lowest for the smallest $k$-bins, and for most bins the FFTLog-based method agrees with direct integration within $0.1\%$. 
To illustrate this point in more detail, Figure~\ref{fig:comparison_T2211_kmax0.4} shows the comparison of the $C^{T_{2211}}$ contribution, which is what our method and direct integration actually compute, on the same $k$-bins as Figure~\ref{fig:comparison_fftlog_vs_direct_kmax0.4}.
We can see that our method computes the $T_{2211}$ contributions within $0.001 \%$ accuracy for most combinations of multipoles.
Relatively larger errors in Figure~\ref{fig:comparison_fftlog_vs_direct_kmax0.4}, especially when a small $k$-bin is involved, are due to large cancellations between $C^{T_{2211}}$ and $C^{T_{3111}}$ contributions.
The errors in the computation increase if we reduce the number of samples, i.e., $N$ in Eq.~(\ref{eq:decomp_pk_lin}), used in the FFTLog operation. 
We show the case of $N = 256$ in Appendix~\ref{sec:fft_res}.

\subsection{Computation time}

Under the setting of the FFTLog-based decomposition and for the 40 $k$-bins we use in the demonstration, it takes about 26 seconds on a 2.8 GHz quad-core Intel Core i7 processor to compute all 23 master integrals, without any explicit parallelization or acceleration.
This is the dominant bottleneck of the whole computation, and the subsequent steps to compute the covariance matrices for all six combinations of multipoles take only less than 0.1 seconds in total (except for the time on running the Boltzmann code to obtain the linear power spectrum).
Since the master integrals do not depend on cosmology, redshift, and galaxy bias parameters, all of the steps affected by these parameters are done very quickly.
This result means that iterative updates of the covariance matrix by the best-fit parameters in cosmology inference can be done with very negligible computation time.

Furthermore, the non-Gaussian part of the covariance of the cross-power spectrum, or of the cross-covariance of power spectra between multiple tracers is described by the cross-trispectrum \cite{Smith:2008ut}.
Since the cross-trispectrum has the same functional form as the auto-trispectrum except for the combination of galaxy bias parameters, we can use the same master integrals as the auto-covariance to compute the non-Gaussian part of multi-tracer covariances.
Hence, our method can be used to compute all necessary covariance matrices for an arbitrary number of tracers in $\O(10)$ seconds (except for the Boltzmann code), starting from the computation of the master integrals.
In contrast, numerical integration methods directly integrate the whole integrand of $T_{2211}$ term in Eq.~(\ref{eq:cov_T0_detail}), requiring one to redo all computation for different cosmology, redshift, galaxy bias parameters, different combinations of multipole moments, and different combinations of tracers.
While our method is already faster than direct numerical integration for a single combination of multipoles of a single tracer, it is expected to become much more advantageous in multi-tracer analysis with multiple redshift bins.

\section{Summary and Conclusion}
\label{sec:conclusion}

The computation of the covariance matrix for the galaxy power spectrum is an essential ingredient for accurate cosmology inference from spectroscopic surveys.
While the simulation-based covariance has been the \textit{de facto} standard in recent analyses of galaxy clustering, we focus on the analytic covariance of the redshift-space power spectrum multipoles, and its computational bottleneck, the non-Gaussian term coming from the connected four-point correlation, i.e., trispectrum.

We propose a fast method to compute the non-Gaussian covariance of the power spectrum multipoles by bypassing the angular integral between two wavevectors $\bk_1$ and $\bk_2$.
It employs the power-law decomposition of the linear power spectrum with the FFTLog so that we can apply the analytic solution of the integral in the problem. 
It is the same approach in spirit as the existing methods proposed in a series of papers \cite{Assassi:2017lea,Simonovic:2017mhp,Schoneberg:2018fis,Lee:2020ebj}, while the purpose is different.
This paper shows the currently widely used FFTLog-based method is also applicable to the non-Gaussian covariance of the redshift-space power spectrum multipoles.

Our method is demonstrated to give accurate computation of the non-Gaussian covariance for all combinations of the power spectrum monopole, quadrupole, and hexadecapole, up to $k_\mathrm{max} \approx 0.4\,h\,\mathrm{Mpc}^{-1}$, i.e., beyond the scales where the perturbation theory-based modeling of the redshift-space galaxy power spectrum is valid.
To complete the computation of all of them, it takes a few 10 seconds on a laptop computer.
More importantly, the computation of the covariance for different values of cosmology, redshift, and galaxy bias parameters is done with negligible computation time ($\sim \O(0.1)$ seconds) once the first computation of the master integrals is completed.
This feature can be highly useful in a situation where we need to compute the covariance matrices for multiple species of tracers at multiple redshifts.

The method in this paper enables computing an integral of the form Eq.~(\ref{eq:integral_element}). 
Therefore, the same or a similar technique could be applicable to computing the covariance of higher-order statistics, most primarily the galaxy bispectrum.
In Ref.~\cite{Sugiyama:2019ike}, the angular integrals appearing in the covariance of the bispectrum multipoles or the cross-covariance of the power spectrum and the bispectrum are done using multidimensional numerical integration.
Combining the analytic integral and FFTLog-based decomposition could be possible to resolve the costly integration and realize fast computation of the covariance of the higher-order statistics, which will be important to unlock cosmological information beyond the power spectrum in ongoing and future galaxy surveys.
We leave this extension for future work.

The numerical code we implemented in this work is publicly available in Ref.~\footnote{\url{https://github.com/archaeo-pteryx/PowerSpecCovFFT}}.

\acknowledgments
The author would like to thank
Xiao~Fang
and Elisabeth~Krause
for useful discussions or detailed comments on this manuscript.
We acknowledge support from the SPHEREx project under a contract from the NASA/GODDARD Space Flight Center to the California Institute of Technology.
This research was also supported by the David and Lucile Packard Foundation.

\appendix

\section{PERTURBATION THEORY KERNELS}
\label{sec:pt_kernel}

The tree-level trispectrum is comprised of perturbations of up to the third order.
In this paper, we calculate the redshift-space galaxy trispectrum based on the following galaxy bias expansion:
\begin{align}
    \delta_{\mathrm{g}} = b_1 \delta + \frac{b_2}{2} \delta^2 + b_{\G_2} \G_2 + \frac{b_3}{6} \delta^3 + b_{\G_3} \G_3 + b_{\delta \G_2} \delta \G_2 + b_{\Gamma_3} \Gamma_3,
\end{align}
and the redshift-space distortions expanded to the third order.
The above expansion is specified by a set of galaxy bias parameters $\{b_1, b_2, b_{\G_2}, b_3, b_{\G_3}, b_{\delta \G_2}, b_{\Gamma_3} \}$. 
$\delta^n$ is the $n$th power of the matter density fluctuations, and we use the Galileon operators \cite{Nicolis:2008in,Eggemeier:2018qae}
\begin{align}
    \G_2(\Phi_\mathrm{g}) &\equiv (\nabla_i \nabla_j \Phi_\mathrm{g})^2 - (\nabla^2 \Phi_\mathrm{g})^2, \\ 
    \G_3(\Phi_\mathrm{g}) &\equiv 2 \nabla_i \nabla_j \Phi_\mathrm{g} \nabla^j \nabla_k \Phi_\mathrm{g} \nabla^k \nabla^i \Phi_\mathrm{g} + (\nabla^2 \Phi_\mathrm{g})^3 - 3 (\nabla_i \nabla_j \Phi_\mathrm{g})^2 \nabla^2 \Phi_\mathrm{g}, \\
    \Gamma_3(\Phi_\mathrm{g}, \Phi_\mathrm{v}) &\equiv \G_2(\Phi_\mathrm{g}) - \G_2(\Phi_\mathrm{v}),
\end{align}
where $\Phi_\mathrm{g}$ and $\Phi_\mathrm{v}$ are the rescaled gravitational and velocity potentials.
The redshift-space density fluctuations of biased tracers in Fourier space are expanded as
\begin{align}
    \delta_\mathrm{g}^{\mathrm{(S)}}(\bk) &= \sum_{n=1}^{\infty} \int_{\bk_1,\cdots,\bk_n} (2\pi)^3 \delta_\mathrm{D}(\bk-\bk_{1 \cdots n}) Z_n(\bk_1, \cdots, \bk_n) \delta^{(1)}(\bk_1) \cdots \delta^{(1)}(\bk_n),
\end{align}
where we use the notation $\int_{\bk_i} \equiv \int \frac{\mathrm{d}^3k_i}{(2\pi)^3}$ and $\bk_{1 \cdots n} \equiv \bk_1+ \cdots +\bk_n$, $\delta_\mathrm{D}(\cdot)$ is the Dirac's delta function, and $\delta^{(1)}(\bk)$ is the linear matter density perturbation.
According to the above bias expansion, the explicit formulas of the PT kernels up to the third order are
\begin{align}
    Z_1(\bk_1) &= b_1 + f \mu_1^2, \\
    Z_2(\bk_1,\bk_2) &= b_1 F_2(\bk_1,\bk_2) + \frac{b_2}{2} + b_{\G_2} \K_{\G_2}(\bk_1,\bk_2) + f \mu_{12}^2 G_2(\bk_1,\bk_2) + \frac{f \mu_{12} k_{12}}{2} b_1 \left( \frac{\mu_1}{k_1} + \frac{\mu_2}{k_2} \right) + \frac{(f \mu_{12} k_{12})^2}{2} \frac{\mu_1}{k_1} \frac{\mu_2}{k_2}, \\
    Z_3(\bk_1,\bk_2,\bk_3) &= b_1 F_3(\bk_1,\bk_2,\bk_3) + b_2 F_2(\bk_2,\bk_3) + 2 b_{\G_2} \K_{\G_2}(\bk_1,\bk_2+\bk_3) F_2(\bk_2,\bk_3) + \frac{b_3}{6} + b_{\delta \G_2} \K_{\G_2}(\bk_2,\bk_3) \nonumber\\
    &\,\, + b_{\G_3} \K_{\G_3}(\bk_1,\bk_2,\bk_3) + 2 b_{\Gamma_3} \K_{\G_2}(\bk_1,\bk_2+\bk_3) [F_2(\bk_2,\bk_3) - G_2(\bk_2,\bk_3)] + f \mu_{123}^2 G_3(\bk_1,\bk_2,\bk_3) \nonumber\\
    &\,\, + (f \mu_{123} k_{123}) b_1 \frac{\mu_{23}}{k_{23}} G_2(\bk_2,\bk_3) + (f \mu_{123} k_{123}) \frac{\mu_1}{k_1} \left[ b_1 F_2(\bk_2,\bk_3) + \frac{b_2}{2} + b_{\G_2} \K_{\G_2}(\bk_2,\bk_3) \right] \nonumber\\
    &\,\, + (f \mu_{123} k_{123})^2 \frac{\mu_1}{k_1} \frac{\mu_{23}}{k_{23}} G_2(\bk_2,\bk_3) + \frac{(f \mu_{123} k_{123})^2}{2} b_1 \frac{\mu_1}{k_1} \frac{\mu_2}{k_2} + \frac{(f \mu_{123} k_{123})^3}{6} \frac{\mu_1}{k_1} \frac{\mu_2}{k_2}  \frac{\mu_3}{k_3},
\end{align}
where $\mu_i \equiv (\bk_i \cdot \hat{\bn}) / k_i$ and $\mu_{1 \cdots n} \equiv (\bk_{1 \cdots n} \cdot \hat{\bn}) / k_{1 \cdots n}$, with $\hat{\bn}$ being the unit vector along the line-of-sight direction.
$F_2$ and $F_3$ are the second- and third-order symmetrized kernels of the density perturbation, and $G_2$ and $G_3$ are those of the velocity (divergence) perturbation.
Note that the $Z_3$ expression above is not symmetrized.
The kernels for the Galileon operators $\G_2, \G_3$ are
\begin{align}
    \mathcal{K}_{\G_2}(\bk_1,\bk_2) &= \frac{(\bk_1\cdot\bk_2)^2}{k_1^2 k_2^2} - 1, \\
    \mathcal{K}_{\G_3}(\bk_1,\bk_2,\bk_3) &= 2 \frac{(\bk_1 \cdot \bk_2) (\bk_2 \cdot \bk_3) (\bk_3 \cdot \bk_1)}{k_1^2 k_2^2 k_3^2} - \left[ \frac{(\bk_1 \cdot \bk_2)^2}{k_1^2 k_2^2} + \frac{(\bk_2 \cdot \bk_3)^2}{k_2^2 k_3^2} + \frac{(\bk_3 \cdot \bk_1)^2}{k_3^2 k_1^2} \right] + 1.
\end{align}
While we can in general calculate the $C^{T_{3111}}$ contribution using the $Z_3$ kernel with nonzero values of the third-order bias parameters $\{ b_3, b_{\G_3}, b_{\delta \G_2}, b_{\Gamma_3} \}$, we set them to be zero in Sec.~\ref{sec:results} for simplicity.

\section{IMPLEMENTATION OF MASTER INTEGRALS}
\label{sec:master_integral}

The master integral Eq.~(\ref{eq:master_integral}) we use in this paper has the hypergeometric function:
\begin{align}
\label{eq:hyp2f1}
    {_2F_1}(a,b;c;z) &= \frac{\Gamma(c)}{\Gamma(a) \Gamma(b)} \sum_{n=0}^{\infty} \frac{
    \Gamma(a+n) \Gamma(b+n)}{\Gamma(c+n)} \frac{z^n}{n!},
\end{align}
where $\Gamma(\cdot)$ is the Gamma function, and the parameters $\{a,b,c\}$ and the argument $z$ are complex-valued in general.
The hypergeometric function is hard to be computed efficiently for a large array of inputs, because the convergence of the series in Eq.~(\ref{eq:hyp2f1}) becomes very slow when $|z|$ is large, especially in the case that it has complex-valued parameters of large absolute values.
Note that, since $|z|$ equals to $2 k_1 k_2 / (k_1^2 + k_2^2)$ in our problem, $|z|$ close to 1 corresponds to the case that $k_1$ and $k_2$ are close to each other and $|z|=1$ corresponds to $k_1 = k_2$.
However, only the case that the parameter $b$ in Eq.~(\ref{eq:hyp2f1}) is a non-negative integer is actually relevant to our problem: $\beta$ in Eq.~(\ref{eq:integral_element}) takes the values $0, \cdots, 12$ for the power spectrum monopole, quadrupole, and hexadecapole, as shown in Table~\ref{tab:combination_ab}.

When $b$ is a non-negative integer, ${_2F_1}\left(a, b+1; b+2; z \right)$ reduces to an elementary function of $z$ and $a$.
More specifically, it has the following closed-form expression:
\begin{align}
\label{eq:hyp2f1_ab}
    {_2F_1}(a,b+1;b+2;z) &= \frac{(-1)^{b+1}(b+1)!}{\prod_{n=1}^{b+1} (a-n) z^{b+1}} \left \{ 1 + \frac{P_{b+1}(z;a)}{(1-z)^a} \right \},
\end{align}
where $P_{n}(z;a)$ is the $n$th-order polynomial of $z$ whose coefficients are polynomials of $a$.
Accordingly, the main part of the master integral is obtained by computing
\begin{align}
\label{eq:hyp2f1_ab_master}
    {_2F_1}(a,b+1;b+2;-z) + (-1)^{b} {_2F_1}(a,b+1;b+2;z) 
    &= \frac{(b+1)!}{\prod_{n=1}^{b+1} (a-n) z^{b+1}} \left \{ \frac{P_{b+1}(-z;a)}{(1+z)^a} - \frac{P_{b+1}(z;a)}{(1-z)^a} \right \}.
\end{align}
This form allows us to avoid direct computation of the hypergeometric function, and the implementation of it gives accurate results when $z$ is not close to zero.
When $z$ is small, both of the two terms in the curly bracket of the above formula become very close to 1, because of the following reason: the $n$th-order polynomial $P_{n}(z;a)$ is equal to the Taylor series of $-(1-z)^a$ up to the $(n-1)$th order, i.e.,
\begin{align}
    (1-z)^a &= - P_{n}(z;a) + \O(z^{n}).
\end{align}
This implies that $P_{n}(z;a) / (1-z)^a$ or $P_{n}(-z;a) / (1+z)^a$ deviate from 1 only by $\O(z^{n})$ when $z$ is close to zero.
Hence, a naive implementation of Eq.~(\ref{eq:hyp2f1_ab_master}) in finite-precision arithmetic leads to a large loss of significance when $z$ is close to zero, which results in large errors for the computed covariance of higher multipoles when $k_1$ and $k_2$ are distant from each other.

Thus, we use the property of the hypergeometric functions that they converge quickly when $|z|$ is small. We approximate the left-hand side of Eq.~(\ref{eq:hyp2f1_ab_master}) by truncating the series Eq.~(\ref{eq:hyp2f1}) at the 20th order for $z \le 0.1$, and evaluate the right-hand side of Eq.~(\ref{eq:hyp2f1_ab_master}) for $0.1 < z < 1$.
This switching technique is similar to that used in Ref.~\cite{Assassi:2017lea} (see its Appendices~A and B).
Finally, the last term of Eq.~(\ref{eq:hyp2f1_ab_master}) diverges for $z=1$. 
However, ${_2F_1}(a,b+1;b+2;-1) + (-1)^{b} {_2F_1}(a,b+1;b+2;1)$ also reduces to a simple function of $a$, for each non-negative integer value of $b$. 

\section{SHOT-NOISE CONTRIBUTION OF NON-GAUSSIAN COVARIANCE}
\label{sec:shotnoise}

In the main text, we focus on the trispectrum terms without the shot-noise contribution arising from the discreteness of the tracers.
The shot-noise contribution of the non-Gaussian covariance is (see e.g. Refs.~\cite{Sugiyama:2019ike,Wadekar:2019rdu})
\begin{align}
\label{eq:cov_sn_connected}
    C^{T_0,\mathrm{SN}}_{\ell_1,\ell_2}(k_1,k_2) &= \frac{1}{V} \frac{1}{N_{k_1}} \sum_{\bk'_1 \in \bin~1} \frac{1}{N_{k_2}} \sum_{\bk'_2 \in \bin~2} (2\ell_1+1) \L_{\ell_1}(\hat{\bk}'_1 \cdot \hat{\bn}) (2\ell_2+1) \L_{\ell_2}(\hat{\bk}'_2 \cdot \hat{\bn}) \nonumber\\
    &\,\, \times \left\{ \frac{4}{\bar{n}} B(\bk'_1,\bk'_2,-\bk'_1-\bk'_2) + \frac{2}{\bar{n}^2} P(\bk'_1+\bk'_2) \right\} \nonumber\\
    &\approx \frac{1}{V} \int \frac{\mathrm{d}^2 \Omega_{\hat{\bk}_1}}{4\pi} \int \frac{\mathrm{d}^2 \Omega_{\hat{\bk}_2}}{4\pi} (2\ell_1+1) \L_{\ell_1}(\hat{\bk}_1 \cdot \hat{\bn}) (2\ell_2+1) \L_{\ell_2}(\hat{\bk}_2 \cdot \hat{\bn}) \left\{ \frac{4}{\bar{n}} B(\bk_1,\bk_2,-\bk_1-\bk_2) + \frac{2}{\bar{n}^2} P(\bk_1+\bk_2) \right\},
\end{align}
where $P$ and $B$ are the redshift-space galaxy power spectrum and bispectrum, and $\bar{n}$ is the mean galaxy number density.
Using the tree-level SPT to model the galaxy power spectrum and bispectrum, i.e.,
\begin{align}
    P(\bk) &= Z_1^2(\bk) P_{\lin}(k), \\
    B(\bk_1,\bk_2,\bk_3) &= 2 Z_1(\bk_1) Z_1(\bk_2) Z_2(\bk_1,\bk_2) P_{\lin}(k_1) P_{\lin}(k_2) + \text{(2 perms.)},
\end{align}
Equation~(\ref{eq:cov_sn_connected}) becomes
\begin{align}
\label{eq:cov_sn_connected_detail}
    C^{T_0,\mathrm{SN}}_{\ell_1,\ell_2}(k_1,k_2) 
    &= \frac{1}{V} \int_{\hat{\bk}_{\ell_1}, \hat{\bk}_{\ell_2}} \frac{8}{\bar{n}} P_{\lin}(k_1) P_{\lin}(k_2) Z_1(\bk_1) Z_1(\bk_2) Z_2(\bk_1,\bk_2) \nonumber\\
    &\,\, + \frac{1}{V} \int_{\hat{\bk}_{\ell_1}, \hat{\bk}_{\ell_2}} \frac{8}{\bar{n}} P_{\lin}(|\bk_1+\bk_2|) Z_1(\bk_1+\bk_2) \left[ P_{\lin}(k_1) Z_1(\bk_1) Z_1(-\bk_1,\bk_1+\bk_2) + (\bk_1 \leftrightarrow \bk_2) \right] \nonumber\\
    &\,\, + \frac{1}{V} \int_{\hat{\bk}_{\ell_1}, \hat{\bk}_{\ell_2}} \frac{2}{\bar{n}^2} P_{\lin}(|\bk_1+\bk_2|) Z_1^2(\bk_1+\bk_2),
\end{align}
where we use the shorthand notation Eq.~(\ref{eq:brev_notation}).
The above formula reduces to the same structure of integral as Eq.~(\ref{eq:cov_T0_int_reduction}), and thus the FFTLog-based method described in this paper is applicable.

\begin{figure}
\centering
  \begin{tabular}{c}
    \begin{minipage}{\hsize}
    \centering
    \includegraphics[width=1\columnwidth]{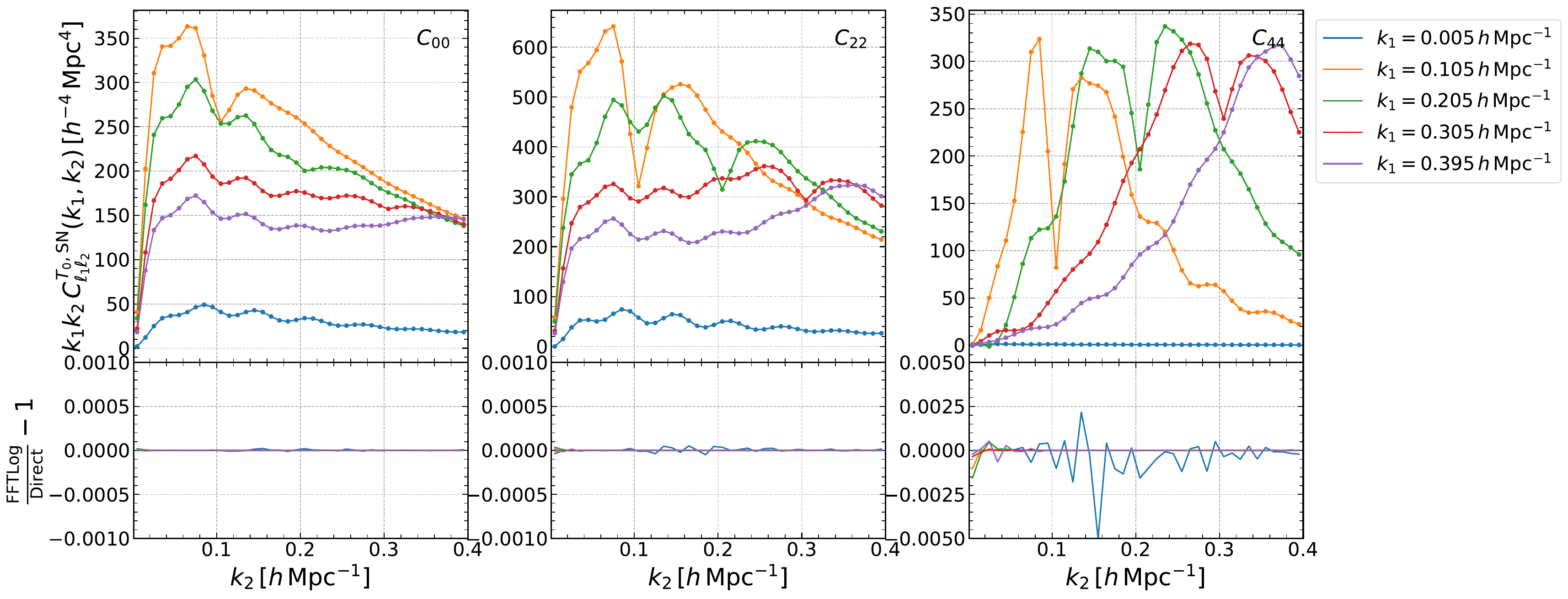}
    \end{minipage} \\
    
    \begin{minipage}{\hsize}
    \centering
    \includegraphics[width=1\columnwidth]{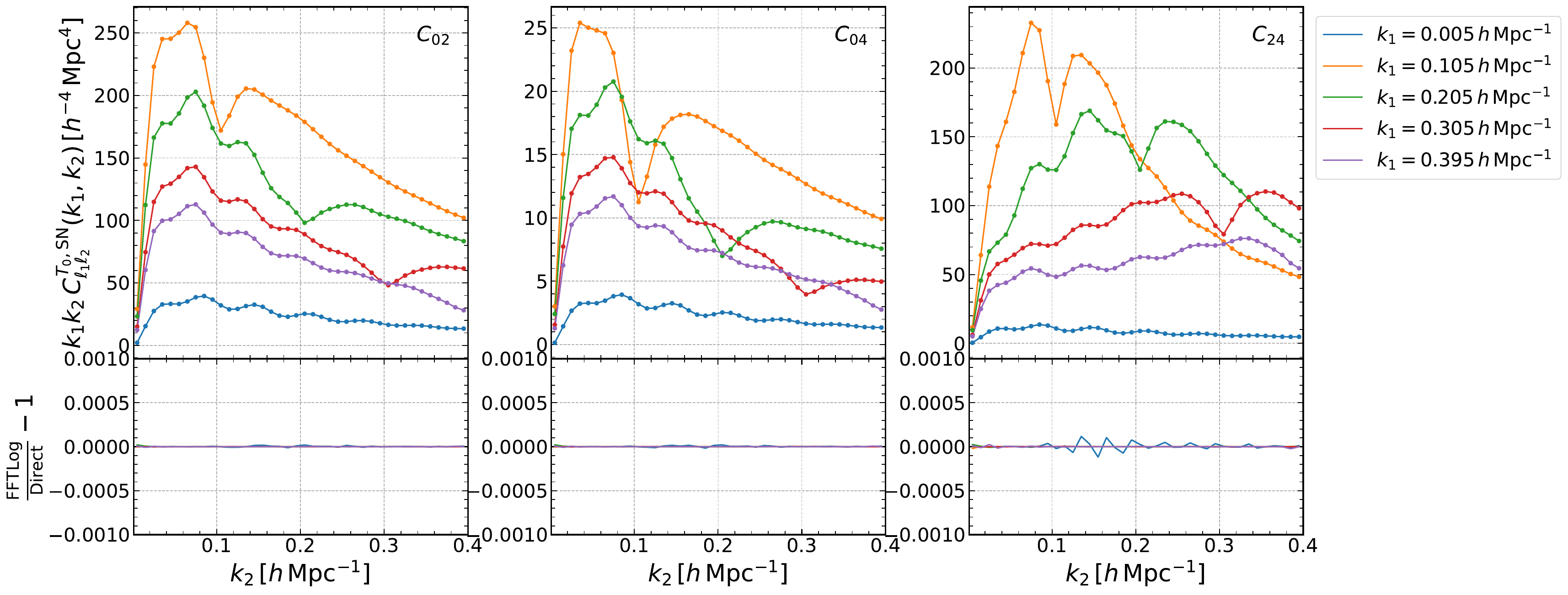}
    \end{minipage}
  \end{tabular}
\caption{
Comparison of the shot-noise contribution of the non-Gaussian covariance, between the FFTLog-based method and direct numerical integration.
}
\label{fig:comparison_T0_SN_kmax0.4}
\end{figure}

Figure~\ref{fig:comparison_T0_SN_kmax0.4} shows the comparison of the shot-noise contribution Eq.~(\ref{eq:cov_sn_connected_detail}) between the FFTLog-based method and direct numerical integration.
In this figure, we set the galaxy number density $\bar{n} = 10^{-4}\,(h^{-1}\,\mathrm{Mpc})^{-3}$ and keep the other parameters, i.e., cosmology, redshift, galaxy bias parameters, and the volume, the same as in Sec.~\ref{sec:results}.
The FFTLog-based method agrees with direct integration within $0.1-0.5\%$.

\section{IMPACT OF FFTLog GRID RESOLUTION}
\label{sec:fft_res}

\begin{figure}
\centering
  \begin{tabular}{c}
    \begin{minipage}{\hsize}
    \centering
    \includegraphics[width=1\columnwidth]{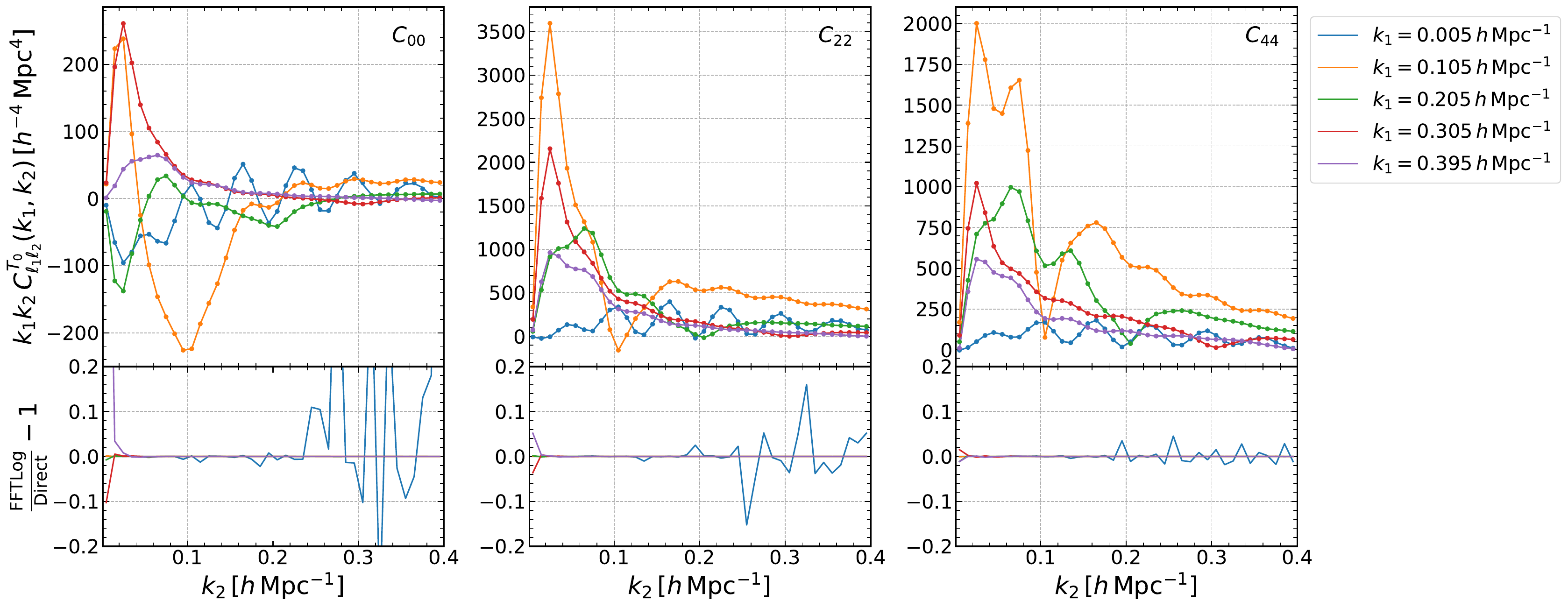}
    \end{minipage} \\
    
    \begin{minipage}{\hsize}
    \centering
    \includegraphics[width=1\columnwidth]{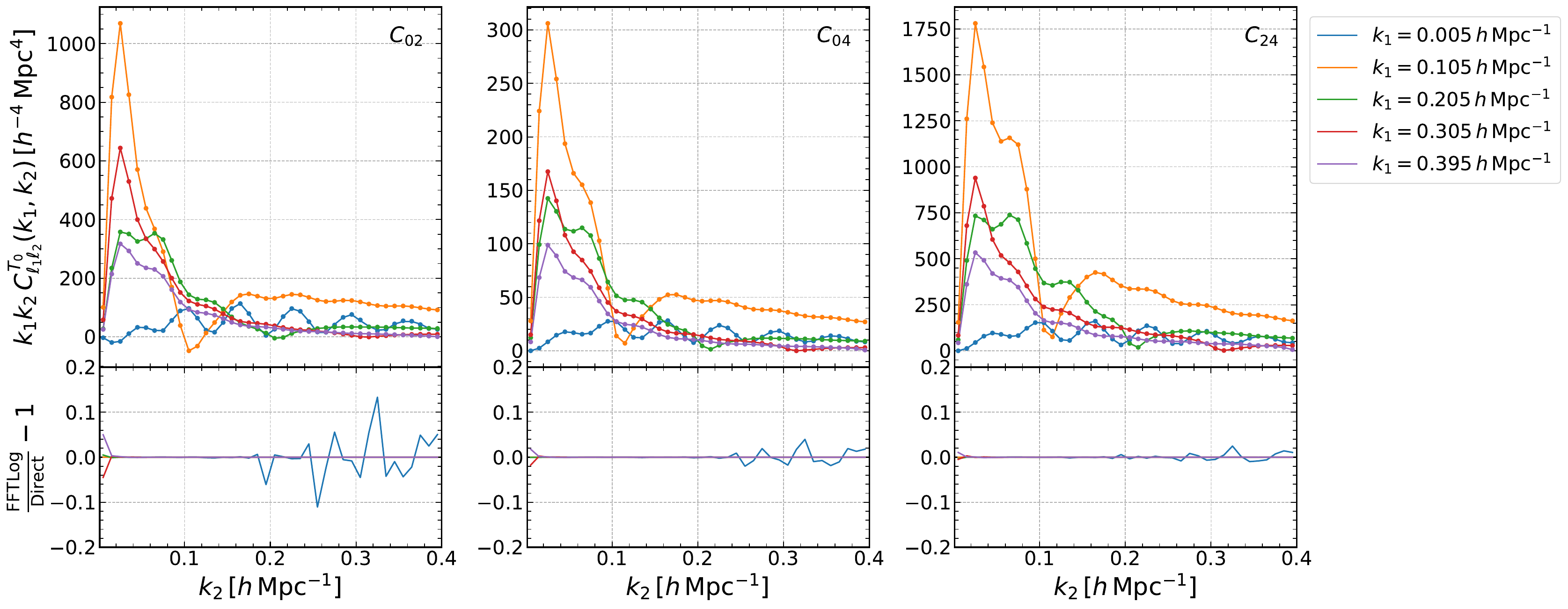}
    \end{minipage}
  \end{tabular}
\caption{
The same as Fig.~\ref{fig:comparison_fftlog_vs_direct_kmax0.4}, but the case we perform the FFTLog-based decomposition of the linear power spectrum with a lower resolution $N = 256$.
In all cases, the FFTLog-based method agrees with direct integration well within $1\%$, except for the case involving the smallest bin of $k (= 0.005\,h\,\mathrm{Mpc}^{-1})$.  
}
\label{fig:comparison_fftlog_vs_direct_kmax0.4_N256}
\end{figure}

Since the FFTLog-based power-law decomposition is an approximation of the linear power spectrum, the accuracy of the resultant covariance matrix depends on the resolution of the logarithmic FFT grid.
This appendix is devoted to showing how the computation accuracy decreases when the FFT grid resolution is decreased.
In the case of $N = 256$, the whole computation time is about 13 seconds (and again, most of the time is for the computation of the master integrals).
Figure~\ref{fig:comparison_fftlog_vs_direct_kmax0.4_N256} shows the comparison of the same covariance as in Fig.~\ref{fig:comparison_fftlog_vs_direct_kmax0.4}, but the case of $N = 256$.
Due to large cancellations between the $C^{T_{2211}}$ and $C^{T_{3111}}$ contributions, the computation accuracy for the bins involving $k = 0.005 \, h\, \mathrm{Mpc}^{-1}$ is somewhat worse.
However, we do not expect this level of difference to affect cosmology inference in practice.

\bibliographystyle{unsrt}
\bibliography{ref}

\begin{thebibliography}{10}

\bibitem{DAmico:2019fhj}
Guido D'Amico, J\'er\^ome Gleyzes, Nickolas Kokron, Katarina Markovic, Leonardo
  Senatore, Pierre Zhang, Florian Beutler, and H\'ector Gil-Mar\'\i{}n.
\newblock {The Cosmological Analysis of the SDSS/BOSS data from the Effective
  Field Theory of Large-Scale Structure}.
\newblock {\em JCAP}, 05:005, 2020.

\bibitem{Ivanov:2019pdj}
Mikhail~M. Ivanov, Marko Simonovi\'c, and Matias Zaldarriaga.
\newblock {Cosmological Parameters from the BOSS Galaxy Power Spectrum}.
\newblock {\em JCAP}, 05:042, 2020.

\bibitem{Chen:2021wdi}
Shi-Fan Chen, Zvonimir Vlah, and Martin White.
\newblock {A new analysis of galaxy 2-point functions in the BOSS survey,
  including full-shape information and post-reconstruction BAO}.
\newblock {\em JCAP}, 02(02):008, 2022.

\bibitem{Philcox:2021kcw}
Oliver H.~E. Philcox and Mikhail~M. Ivanov.
\newblock {BOSS DR12 full-shape cosmology: \ensuremath{\Lambda}CDM constraints
  from the large-scale galaxy power spectrum and bispectrum monopole}.
\newblock {\em Phys. Rev. D}, 105(4):043517, 2022.

\bibitem{Kobayashi:2021oud}
Yosuke Kobayashi, Takahiro Nishimichi, Masahiro Takada, and Hironao Miyatake.
\newblock {Full-shape cosmology analysis of the SDSS-III BOSS galaxy power
  spectrum using an emulator-based halo model: A 5\% determination of
  \ensuremath{\sigma}8}.
\newblock {\em Phys. Rev. D}, 105(8):083517, 2022.

\bibitem{DESI:2016fyo}
Amir Aghamousa et~al.
\newblock {The DESI Experiment Part I: Science,Targeting, and Survey Design}.
\newblock 10 2016.

\bibitem{EUCLID:2011zbd}
R.~Laureijs et~al.
\newblock {Euclid Definition Study Report}.
\newblock 10 2011.

\bibitem{Dore:2014cca}
Olivier Dor\'e et~al.
\newblock {Cosmology with the SPHEREX All-Sky Spectral Survey}.
\newblock 12 2014.

\bibitem{PFSTeam:2012fqu}
Masahiro Takada et~al.
\newblock {Extragalactic science, cosmology, and Galactic archaeology with the
  Subaru Prime Focus Spectrograph}.
\newblock {\em Publ. Astron. Soc. Jap.}, 66(1):R1, 2014.

\bibitem{Gehrels:2014spa}
Neil Gehrels and David~N. Spergel.
\newblock {Wide-Field InfraRed Survey Telescope (WFIRST) Mission and Synergies
  with LISA and LIGO-Virgo}.
\newblock {\em J. Phys. Conf. Ser.}, 610(1):012007, 2015.

\bibitem{Chuang:2014vfa}
Chia-Hsun Chuang, Francisco-Shu Kitaura, Francisco Prada, Cheng Zhao, and
  Gustavo Yepes.
\newblock {EZmocks: extending the Zel'dovich approximation to generate mock
  galaxy catalogues with accurate clustering statistics}.
\newblock {\em Mon. Not. Roy. Astron. Soc.}, 446:2621--2628, 2015.

\bibitem{Kitaura:2015uqa}
Francisco-Shu Kitaura et~al.
\newblock {The clustering of galaxies in the SDSS-III Baryon Oscillation
  Spectroscopic Survey: mock galaxy catalogues for the BOSS Final Data
  Release}.
\newblock {\em Mon. Not. Roy. Astron. Soc.}, 456(4):4156--4173, 2016.

\bibitem{Rodriguez-Torres:2015vqa}
Sergio~A. Rodr\'\i{}guez-Torres et~al.
\newblock {The clustering of galaxies in the SDSS-III Baryon Oscillation
  Spectroscopic Survey: modelling the clustering and halo occupation
  distribution of BOSS CMASS galaxies in the Final Data Release}.
\newblock {\em Mon. Not. Roy. Astron. Soc.}, 460(2):1173--1187, 2016.

\bibitem{Balaguera-Antolinez:2022xko}
Andr\'es Balaguera-Antol\'\i{}nez et~al.
\newblock {DESI Mock Challenge: Halo and galaxy catalogs with the bias
  assignment method}.
\newblock {\em Astron. Astrophys.}, 673:A130, 2023.

\bibitem{Hartlap:2006kj}
J.~Hartlap, Patrick Simon, and P.~Schneider.
\newblock {Why your model parameter confidences might be too optimistic:
  Unbiased estimation of the inverse covariance matrix}.
\newblock {\em Astron. Astrophys.}, 464:399, 2007.

\bibitem{Dodelson:2013uaa}
Scott Dodelson and Michael~D. Schneider.
\newblock {The Effect of Covariance Estimator Error on Cosmological Parameter
  Constraints}.
\newblock {\em Phys. Rev. D}, 88:063537, 2013.

\bibitem{Percival:2013sga}
Will~J. Percival et~al.
\newblock {The Clustering of Galaxies in the SDSS-III Baryon Oscillation
  Spectroscopic Survey: Including covariance matrix errors}.
\newblock {\em Mon. Not. Roy. Astron. Soc.}, 439(3):2531--2541, 2014.

\bibitem{Meiksin:1998mu}
A.~Meiksin and Martin~J. White.
\newblock {The Growth of correlations in the matter power spectrum}.
\newblock {\em Mon. Not. Roy. Astron. Soc.}, 308:1179, 1999.

\bibitem{Scoccimarro:1999kp}
Roman Scoccimarro, Matias Zaldarriaga, and Lam Hui.
\newblock {Power spectrum correlations induced by nonlinear clustering}.
\newblock {\em Astrophys. J.}, 527:1, 1999.

\bibitem{Smith:2008ut}
Robert~E. Smith.
\newblock {Covariance of cross-correlations: towards efficient measures for
  large-scale structure}.
\newblock {\em Mon. Not. Roy. Astron. Soc.}, 400:851, 2009.

\bibitem{Chan:2016ehg}
Kwan~Chuen Chan and Linda Blot.
\newblock {Assessment of the Information Content of the Power Spectrum and
  Bispectrum}.
\newblock {\em Phys. Rev. D}, 96(2):023528, 2017.

\bibitem{Sugiyama:2019ike}
Naonori~S. Sugiyama, Shun Saito, Florian Beutler, and Hee-Jong Seo.
\newblock {Perturbation theory approach to predict the covariance matrices of
  the galaxy power spectrum and bispectrum in redshift space}.
\newblock {\em Mon. Not. Roy. Astron. Soc.}, 497(2):1684--1711, 2020.

\bibitem{Wadekar:2019rdu}
Digvijay Wadekar and Roman Scoccimarro.
\newblock {Galaxy power spectrum multipoles covariance in perturbation theory}.
\newblock {\em Phys. Rev. D}, 102(12):123517, 2020.

\bibitem{Wadekar:2020hax}
Digvijay Wadekar, Mikhail~M. Ivanov, and Roman Scoccimarro.
\newblock {Cosmological constraints from BOSS with analytic covariance
  matrices}.
\newblock {\em Phys. Rev. D}, 102:123521, 2020.

\bibitem{BOSS:2012dmf}
Kyle~S. Dawson et~al.
\newblock {The Baryon Oscillation Spectroscopic Survey of SDSS-III}.
\newblock {\em Astron. J.}, 145:10, 2013.

\bibitem{TALMAN197835}
James~D Talman.
\newblock Numerical fourier and bessel transforms in logarithmic variables.
\newblock {\em Journal of Computational Physics}, 29(1):35--48, 1978.

\bibitem{Hamilton:1999uv}
A.~J.~S. Hamilton.
\newblock {Uncorrelated modes of the nonlinear power spectrum}.
\newblock {\em Mon. Not. Roy. Astron. Soc.}, 312:257--284, 2000.

\bibitem{Schmittfull:2016jsw}
Marcel Schmittfull, Zvonimir Vlah, and Patrick McDonald.
\newblock {Fast large scale structure perturbation theory using one-dimensional
  fast Fourier transforms}.
\newblock {\em Phys. Rev. D}, 93(10):103528, 2016.

\bibitem{Schmittfull:2016yqx}
Marcel Schmittfull and Zvonimir Vlah.
\newblock {FFT-PT: Reducing the two-loop large-scale structure power spectrum
  to low-dimensional radial integrals}.
\newblock {\em Phys. Rev. D}, 94(10):103530, 2016.

\bibitem{McEwen:2016fjn}
Joseph~E. McEwen, Xiao Fang, Christopher~M. Hirata, and Jonathan~A. Blazek.
\newblock {FAST-PT: a novel algorithm to calculate convolution integrals in
  cosmological perturbation theory}.
\newblock {\em JCAP}, 09:015, 2016.

\bibitem{Fang:2016wcf}
Xiao Fang, Jonathan~A. Blazek, Joseph~E. McEwen, and Christopher~M. Hirata.
\newblock {FAST-PT II: an algorithm to calculate convolution integrals of
  general tensor quantities in cosmological perturbation theory}.
\newblock {\em JCAP}, 02:030, 2017.

\bibitem{Assassi:2017lea}
Valentin Assassi, Marko Simonovi\'c, and Matias Zaldarriaga.
\newblock {Efficient evaluation of angular power spectra and bispectra}.
\newblock {\em JCAP}, 11:054, 2017.

\bibitem{Simonovic:2017mhp}
Marko Simonovi\'c, Tobias Baldauf, Matias Zaldarriaga, John~Joseph Carrasco,
  and Juna~A. Kollmeier.
\newblock {Cosmological perturbation theory using the FFTLog: formalism and
  connection to QFT loop integrals}.
\newblock {\em JCAP}, 04:030, 2018.

\bibitem{Schoneberg:2018fis}
Nils Sch\"oneberg, Marko Simonovi\'c, Julien Lesgourgues, and Matias
  Zaldarriaga.
\newblock {Beyond the traditional Line-of-Sight approach of cosmological
  angular statistics}.
\newblock {\em JCAP}, 10:047, 2018.

\bibitem{Fang:2019xat}
Xiao Fang, Elisabeth Krause, Tim Eifler, and Niall MacCrann.
\newblock {Beyond Limber: Efficient computation of angular power spectra for
  galaxy clustering and weak lensing}.
\newblock {\em JCAP}, 05:010, 2020.

\bibitem{Lee:2020ebj}
Hayden Lee and Cora Dvorkin.
\newblock {Cosmological Angular Trispectra and Non-Gaussian Covariance}.
\newblock {\em JCAP}, 05:044, 2020.

\bibitem{Fang:2020vhc}
Xiao Fang, Tim Eifler, and Elisabeth Krause.
\newblock {2D-FFTLog: Efficient computation of real space covariance matrices
  for galaxy clustering and weak lensing}.
\newblock {\em Mon. Not. Roy. Astron. Soc.}, 497(3):2699--2714, 2020.

\bibitem{Hamilton:2005dx}
Andrew J.~S. Hamilton, Christopher~D. Rimes, and Roman Scoccimarro.
\newblock {On measuring the covariance matrix of the nonlinear power spectrum
  from simulations}.
\newblock {\em Mon. Not. Roy. Astron. Soc.}, 371:1188--1204, 2006.

\bibitem{Takada:2013wfa}
Masahiro Takada and Wayne Hu.
\newblock {Power Spectrum Super-Sample Covariance}.
\newblock {\em Phys. Rev. D}, 87(12):123504, 2013.

\bibitem{Bernardeau:2001qr}
F.~Bernardeau, S.~Colombi, E.~Gaztanaga, and R.~Scoccimarro.
\newblock {Large scale structure of the universe and cosmological perturbation
  theory}.
\newblock {\em Phys. Rept.}, 367:1--248, 2002.

\bibitem{Fry:1983cj}
James~N. Fry.
\newblock {The Galaxy correlation hierarchy in perturbation theory}.
\newblock {\em Astrophys. J.}, 279:499--510, 1984.

\bibitem{Note1}
\protect \url {https://github.com/JayWadekar/CovaPT}.

\bibitem{Lewis:2002ah}
Antony Lewis and Sarah Bridle.
\newblock {Cosmological parameters from CMB and other data: A Monte Carlo
  approach}.
\newblock {\em Phys. Rev. D}, 66:103511, 2002.

\bibitem{Planck:2015fie}
P.~A.~R. Ade et~al.
\newblock {Planck 2015 results. XIII. Cosmological parameters}.
\newblock {\em Astron. Astrophys.}, 594:A13, 2016.

\bibitem{Note2}
\protect \url {https://github.com/archaeo-pteryx/PowerSpecCovFFT}.

\bibitem{Nicolis:2008in}
Alberto Nicolis, Riccardo Rattazzi, and Enrico Trincherini.
\newblock {The Galileon as a local modification of gravity}.
\newblock {\em Phys. Rev. D}, 79:064036, 2009.

\bibitem{Eggemeier:2018qae}
Alexander Eggemeier, Roman Scoccimarro, and Robert~E. Smith.
\newblock {Bias Loop Corrections to the Galaxy Bispectrum}.
\newblock {\em Phys. Rev. D}, 99(12):123514, 2019.

\end{thebibliography}

\end{document}